\documentclass[12pt]{spieman}  
\usepackage{amsmath,amsfonts,amssymb}
\usepackage{graphicx}
\usepackage{setspace}
\usepackage{tocloft}
\usepackage{soul}

\title{Optical ray tracing of echelle spectrographs applied to the wavelength solution for precise radial velocities}

\author[a,b,d]{Marcelo Tala Pinto}
\author[b]{Adrian Kaminski}
\author[b]{Andreas Quirrenbach}
\author[c]{Mathias Zechmeister}
\affil[a]{Facultad de Ingeniería y Ciencias, Universidad Adolfo Ibáñez, Av. Diagonal las Torres 2640, Peñalolén, Santiago, Chile}
\affil[b]{Landessternwarte, Zentrum f\"ur Astronomie der Universit\"at Heidelberg, K\"onigstuhl 12, 69117 Heidelberg, Germany}
\affil[c]{Institut f\"ur Astrophysik und Geophysik, Georg-August-Universit\"at, Friedrich-Hund-Platz 1, 37077 G\"ottingen, Germany}
\affil[d]{Millenium Institute of Astrophysics, Santiago, Chile}

\cftpagenumbersoff{figure}
\cftpagenumbersoff{table} 
\begin{document} 
\maketitle

\begin{abstract}
We present \texttt{moes}, a ray tracing software package that computes the path of rays through echelle spectrographs. Our algorithm is based on sequential direct tracing with Seidel aberration corrections applied at the detector plane.  As a test case, we model the CARMENES VIS spectrograph. After subtracting the best model from the data, the residuals yield an rms of 0.024~pix, setting a new standard to the precision of the wavelength solution of state-of-the-art radial velocity instruments. By including the influence of the changes of the environment in the ray propagation, we are able to predict instrumental radial velocity systematics at the 1~m/s level.
\end{abstract}


{\noindent \footnotesize{*}M. Tala Pinto,  \linkable{marcelo.tala@edu.uai.cl} }

\begin{spacing}{2}   

\section{Introduction}
\label{sect:intro}  
The rapid evolution of astronomical instrumentation in the last decades has enormously increased the standards of quality control for observations. The complexity of state-of-the-art instruments requires full control over the conditions in which observations are performed to optimize their scientific exploitation. This is particularly relevant for echelle spectrographs used in radial velocity (RV) exoplanet searches. This type of instrument requires an extremely high opto-mechanical stability to reach the RV precision required for the detection of substellar companions. In echelle spectrographs with resolution of about 100\,000  typically used on exoplanet searches, 1 pixel in the detector plane corresponds to roughly 1 km/s in velocity space. In order to sense the subtle signals of Earth-like planets, which are of the order of a few m/s or even less, we must sense shifts of the spectra on the detector of the order of a thousandth of a pixel. Hence, most current echelle spectrographs are placed in gravity-invariant, thermally and pressure-controlled environments to optimize their opto-mechanical stability. 

One key element in the measurement of precise RVs is the instrument calibration procedure, as the calibration data serve as a reference for tracking instrumental systematics.
The spectrograph wavelength calibration procedure basically consists in measuring with high accuracy and precision the positions of a set of spectral lines on the detector. With this information one can create a function mapping the detector pixel space to wavelength space in the stellar spectra, the so-called \textit{wavelength solution} (WLS).
The precision of the stellar RVs relies, fundamentally, on the accuracy and precision of the WLS.

Traditionally, astronomical spectrographs are calibrated by using empirical methods, e.g., the wavelength solution is derived by fitting a polynomial function to a sparse calibration line spectrum. Instrument calibrations are performed before the observation night, and the WLS is created based on these calibration data. In the controlled conditions in which echelle spectrographs are operated, it is not expected that their environment will vary considerably between afternoon calibrations and nightly observations. 
With this strategy it is not possible, however, to obtain any information of the internal configuration of the instrument at the time of the observations, i.e., empirical changes of the WLS are not traced back to physical shifts of optical elements within the spectrograph.

Ballester et al. 1997\cite{Ballester97} proposed a physical model description of the instrument to build the WLS. 
Their model contained only the on-axis configuration of the optics, with all aberrations included together as a polynomial at the detector plane. This approach was successfully applied to FOS and STIS on HST \cite{Kerber05}, and to CRIRES and X-shooter on the VLT \cite{Bristow06, Bristow10}. 
Later on, Chanumolu et al. 2015\cite{chan15} developed a physical model for the Hanle echelle spectrograph based on paraxial ray tracing, including exact corrections for some surface types and Buchdahl aberration coefficients for complex modules such as camera systems. This modeling approach was able to track the differences in the centroids between the fiber images across the spectrum, as well as instrumental shifts between two fibers \cite{chanumolu17}.
\begin{figure*}
    \centering
    \includegraphics[width=\textwidth,trim=12 12 0 0,clip]{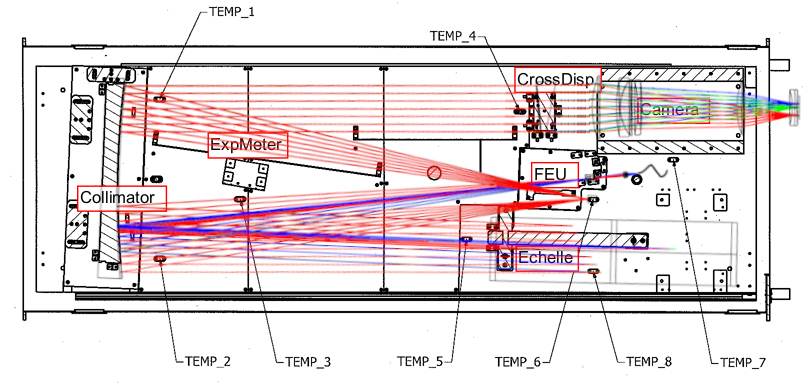}
    \caption{CARMENES VIS optical bench with the optical path of rays through the spectrograph optics overplotted.}
    \label{fig:spec_trace}
\end{figure*}

More recently, we presented a modeling scheme based on direct ray tracing of the chief ray from the slit to the detector plane of an echelle spectrograph, applying exact corrections when tracing through curved surfaces \cite{moes2018}. The direct ray tracing provides an accurate theoretical description of the instrument as it calculates the exact position of each ray along the optical path of the spectrograph. 
Emission line spectra are used to derive the free parameters of the model, which are adjusted to match the observations, thus fully describing the instrument configuration. 

Here we present the {M}odeling {O}ptics of {E}chelle {S}pectrograph software (\texttt{moes}), a ray tracing package that calculates the path of single rays through optical surfaces based on Fermat's principle. \texttt{moes} can model the optical paths of light rays through an echelle spectrograph from the slit to the detector, and use them to build the spectrograph's WLS and to characterize the expected instrumental RV systematics arising from changes in spectrograph's environment. We have applied \texttt{moes} to CARMENES \cite{Quirrenbach16} as a test case. 

CARMENES uses the simultaneous calibration technique to measure precise stellar RVs \cite{baranne96}. This technique consists of observing the stellar spectrum simultaneously with a calibration spectrum, where the latter works as a reference to track any instrumental RV systematics that could arise during the observations.
RV systematics are traced by calculating a Nightly Zero Point (NZP), which corresponds to an instrumental RV offset that is calculated by using all the stars with small RV variability (RV-quiet stars) observed in a given night. These NZPs are calculated for each scientific observation during the night, mitigating the effect of RV systematics that may arise during this period.

Here we propose an environmentally-driven approach to the instrumental systematics, which we call the Temperature-based Zero Point (TZP). By calculating the difference in position of the science and calibration spectra, induced by the changes in the environmental conditions at the time of the observations, we can calculate the TZPs, an RV systematic that can be used to correct the measured stellar RVs and improve their accuracy, only using the information of the temperature and pressure at the time any observation occured.

The paper is organized as follows: In Sect.~\ref{sec:moes} we present the \texttt{moes} ray tracing algorithm, in Sect.~\ref{sec:CARM_VIS} we present the CARMENES VIS \texttt{moes} WLS with the model optimization algorithm, in Sect.~\ref{sec:TZP} we present the computation of the TZP as a model-based RV systematic induced by changes in the spectrograph environment, in Sect.~\ref{sec:discussion} we discuss our results and in Sect.~\ref{sec:Conclusion} we present the conclusions of this research. In the Appendix~\ref{app:environment} we present a description of the thermal expansion model used in our calculations.

\section{The \texttt{MOES} package}\label{sec:moes}
\texttt{moes}\footnote{\href{https://github.com/mtalapinto/moes}{https://github.com/mtalapinto/moes}} is a software package written in python that computes the path of single rays through echelle spectrographs. The basic principles of our ray tracing scheme was presented by \cite{moes2018}. In the CARMENES case the modules that describe the optical elements of the spectrograph are: slit, FN-system (FN stands for F-number, also known as focal ratio), collimator, echelle grating, transfer mirror, cross-dispersion grism, camera and detector. The FN-system is an optical system placed in front of the spectrograph that transforms the focal ratio of the output beam of the fibers from the telescope into the focal ratio of the collimator.

Ray coordinates are described by a position vector $\Vec{X}=[X,Y,Z]$ and a direction vector $\Vec{D}=[d_X, d_Y, d_X]$ in a three-dimensional Euclidean space. The default input to the model are: (a) the initial ray positions and directions in the first surface, which are set to $\Vec{X_0}=[0, 0, 0]$ and $\Vec{D_0}=[0, 0, 1]$, (b) the wavelength $\lambda$, (c) the echelle grating spectral order $o$ of a given spectral feature, (d) a set of instrumental parameters $\Vec{P}$ and (e) a set of environmental parameters $\Vec{T}$. The output is the $(x, y)$ coordinate in the detector plane in pixels and can be written as
\begin{equation}
    (x, y) = \mathrm{moes}(o, \lambda, \Vec{X_0}, \Vec{D_0}, \Vec{P}, \Vec{T})~~.
\end{equation}

The environmental conditions are considered by including a model of the refractive index as a function of wavelength, temperature and pressure, and a model of the thermal expansion of the opto-mechanics (see Appendix~\ref{app:environment} for details). 
The modeling of the refractive index is based on the algorithm used by the commercial optical design software OpticStudio\footnote{\href{https://support.zemax.com/hc/en-us/articles/1500005576002-How-OpticStudio-calculates-refractive-index-at-arbitrary-temperatures-and-pressures}{See web page: How OpticStudio calculates refractive index at arbitrary temperatures and pressure.}}.
The full parameter space of the instrument model consist of 43 opto-mechanical parameters including the positions, orientations and main features of the optical elements that compose the spectrograph plus nine environmental parameters, which represent eight temperature sensors and one pressure sensor placed inside the spectrograph vacuum vessel.

\subsection{Aberration corrections}
Optical aberrations arise from the geometry of the optical surfaces of an imaging system and from changes in the refractive indices of these optical elements and the surrounding medium. In the case of echelle spectrographs, these effects are observed directly in the detector plane. Here we have employed the classical Seidel aberration equations \cite{mahajan14} to model the spectrograph aberrations directly at the detector plane by using a Bayesian interference technique to estimate the best-fit parameters of the aberration model to the residuals of the \texttt{moes} instrument model. We correct for the aberration model by adding the value of the correction {only} to the $x$-coordinate of the spectral lines at the detector plane. 

{Whilst we could correct simultaneously for aberrations along the $y$-axis\cite{chan15}, such correction do not yield significant improvements in the construction of the WLS to the desired accuracy. This could be, most likely, because the WLS represents the relation between the $x$-coordinate and $\lambda$ in the detector plane.}
{Moreover, the $y$-coordinates we use to describe the centroids of the spectral lines have their origin during the processing of the images of the projections of the spectrograph's sliced pseudo-slit in the detector plane. 
The simplicity of our modeling approach might be insufficient to properly model the $y$-coordinate of the spectral lines centroids, without building a suitable model of the PSF for each line. 
Therefore, given that the construction of the instrumental WLS using a physical modeling approach is computationally expensive, we focus only in the $x$-coordinate. This will allow us, at the same time, to optimize the computational resources.}

We have divided aberration corrections into two groups:  mono-chromatic and chromatic. They are calculated during the model optimization procedure. After estimating the best-fit instrumental parameters, we fit the aberration functions to the post-fit residuals of the instrument model.

\subsubsection{Chromatic aberrations}

Chromatic aberrations arise from the wavelength dependence of the refraction angles at optical surfaces. 
In \texttt{moes} the chromatic aberration function is described by
\begin{equation}
     \delta x_{\rm_C}(\lambda) = c_0 \lambda^2+ c_1 + c_2 \lambda^{-2} + c_3 \lambda^{-4}~~,
\end{equation}
where $\delta x_{\rm_C}(\lambda)$ is the coordinate of a point with respect to its position in an image plane free of chromatic aberrations, and $c_0$, $c_1$, $c_2$ and $c_3$ are the \textit{chromatic aberration coefficients}.

\renewcommand{\arraystretch}{1.5}

\begin{table}
    \caption{Summary of the \texttt{moes} parameter space sensitivity calculation for the CARMENES VIS spectrograph. From a total of 45 instrumental parameters, we only show the nine most relevant ones. They are ordered according to their average impact on the $x$-positions of the spectral features in the detector plane. Tilt angles are measured in degrees and distances are measured in mm.}
    \label{tab:error_budget}
    \centering
    \begin{tabular}{@{}l c c c@{}}
      \hline\hline
    Parameter $p$  & $\delta p$ & rms $\delta x_T$~[pix] & rms $\delta y_T$~[pix] \\ \hline
    Echelle constant $G$ [mm$^{-1}$] & 4.62$\cdot$10$^{-6}$ & 4.13 & 0.18 \\
    Collimator x-tilt [deg] & 6.09$\cdot$10$^{-4}$ & 0.59 & 0.02\\
    Echelle incidence angle [deg] & 6.09$\cdot$10$^{-4}$ & 0.52 & 0.02 \\ 
    Grism x-tilt [deg] & 6.09$\cdot$10$^{-4}$ & 0.22 & 2.39$\cdot$10$^{-3}$ \\
    Camera x-tilt [deg] & 6.09$\cdot$10$^{-4}$ & 0.22 & 9.91$\cdot$10$^{-4}$ \\
    Collimator y-tilt [deg] &  6.09$\cdot$10$^{-4}$ & 9.51$\cdot$10$^{-2}$ & 0.46 \\
    Echelle $\gamma$-angle [deg] & 6.09$\cdot$10$^{-4}$ & 4.59$\cdot$10$^{-2}$ & 0.45 \\
    Slit y-decenter [mm]& 6.09$\cdot$10$^{-4}$ & 3.81$\cdot$10$^{-2}$ & 4.85$\cdot$10$^{-3}$ \\
    Field flattener y-decenter [mm] & 6.09$\cdot$10$^{-4}$ & 2.99$\cdot$10$^{-2}$ & 2.42$\cdot$10$^{-4}$ \\
    \hline
    \end{tabular}
\end{table}

\subsubsection{Monochromatic aberrations}

Monochromatic aberrations arise from the shape and geometry of optical surfaces. Hence, the position on the focal plane of a given ray will depend on its position at the optical surfaces. In $\texttt{moes}$ we model the classical Seidel aberrations \cite{mahajan14}. These optical aberrations can be described by
\begin{equation}\label{monab}
\begin{split}
    A_{\rm_O}(\rho, \theta) = a_0\rho\cos \theta + a_1\rho^2 + a_2 \rho^2\cos^2\theta + a_3\rho^3\cos\theta \\+ a_4\rho^3\cos^3\theta + a_5\rho^4 + a_6\rho^4\cos^2\theta + a_7\rho^5\cos\theta + a_8\rho^6~~,
\end{split}
\end{equation}
where $a_{l}$, with $l = 0 \ldots 8$, are the Seidel aberration coefficients. $\rho$ and $\theta$ correspond to the normalized coordinates of a ray in the focal plane in polar coordinates and are given by
\begin{equation}\label{car2pol}
    \rho = \sqrt{x^2 + y^2}~~~{\rm and}~~~
    \theta = \arctan\frac{y}{x}~~.
\end{equation}

Equations~(\ref{monab}) and (\ref{car2pol}) describe wavefront aberrations. Wavefront and ray aberrations are related to each other according to
\begin{equation}\label{rayab}
    \delta x_{\rm_O} = \frac{R}{n}\frac{\partial A_{\rm_O}}{\partial x}~~,
\end{equation}
where $\delta x_{\rm_O}$ is the x-coordinate of a point with respect to its position in an ideal image plane, $R$ is the radius of the reference sphere to which aberrations are referred, and $n$ is the refractive index of the medium. For simplicity we chose $R=1$, so ray aberrations can be computed using Eqs.~(\ref{monab}) and (\ref{rayab}). 

\subsection{Full model}
Finally, the \texttt{moes} x-coordinate consists of three components: a ray tracing component and two aberration terms. This can be written as
\begin{equation}\label{eq:full_moes}
    x = x_{\rm RT}(o, \lambda, \Vec{T}, \Vec{P}) + \delta x_{\rm C}(\lambda) + \delta x_{\rm O}(x, y)~~,
\end{equation}
where $x_{\rm RT}$ is the ray tracing component, and $\delta x_{\rm O}$ and $\delta x_{\rm C}$ are the optical and chromatic aberration components, respectively. We emphasize that to calculate the term $\delta x_{\rm O}$, one must first correct chromatic aberrations.

\begin{figure}
    \centering
    \includegraphics[width=0.85\textwidth]{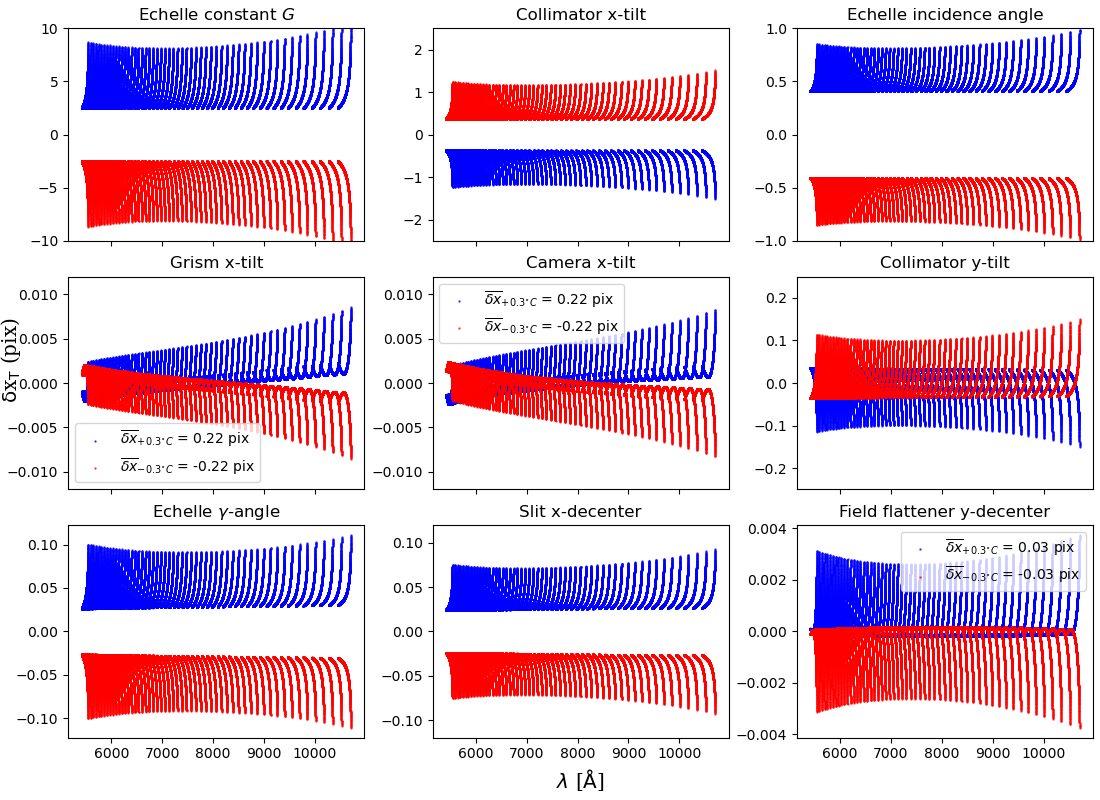}
    \caption{Change in the line positions $\delta x$ due to temperature change $\delta T$ of $+0.3^{\circ}$C (red) and $-0.3^{\circ}$C (blue points) for nine optical elements in the CARMENES VIS model. The grism x-tilt and camera x-tilt points have, additionally, an average decentering of $\mp$0.23~pix due to the $\pm$0.3 temperature change.}
    \label{fig:error_bud_vis}
\end{figure}

\section{CARMENES VIS modeling with  \texttt{moes}}\label{sec:CARM_VIS}

\subsection{CARMENES project}

CARMENES is an instrument installed at the 3.5\,m telescope at the Centro Astronómico Hispano-Alemán (CAHA) on Calar Alto, Spain. It was built with the main science objective of carrying out a precision RV survey of about 300 late-type main-sequence stars to detect low-mass planets in their habitable zones \cite{Quirrenbach16}. CARMENES consists of two echelle spectrographs that observe simultaneously in the visible (VIS) between 5200~-~9600~\AA{} at a spectral resolving power of $\mathcal{R}~=~94\,600$ and in the near-infrared (NIR) between 9600~-~17\,100~\AA{} at $\mathcal{R}=80\,400$. They are fed with starlight from the Cassegrain focus via an optical fiber link \cite{carmenes_fibras} and they are housed in temperature-stabilized vacuum tanks to ensure optimal opto-mechanical stability\cite{carmenes_vis_walter, Becerril2012}.
{The CARMENES VIS spectrograph operates at room temperature ($\sim$283$^{\circ}$K) with small changes within 0.01~K, while pressure is expected to be about 10$^{-5}$mbar\cite{carmenes_vis_walter}.}
The drawing in Fig.~\ref{fig:spec_trace} shows a view of the spectrograph optical bench seen from the top, together with the path of light through the spectrograph optics from the fiber entry to the detector. We have marked the position of the temperature sensors placed on the optical bench that are used to model the environmental effects of the \texttt{moes} WLS.

The wavelength calibration strategy for CARMENES is based on a combination of hollow cathode lamp (HCL) and Fabry-Perot (FP) spectra. The FP is also used simultaneously with science observations to track instrumental drifts during the night. The WLS is computed using a 2D polynomial of the form $x(\lambda, o) = \mathrm{poly}(o\lambda, o)$ in direct regression \cite{Bauer15}.
The stellar RVs are measured from a high signal-to-noise template created by co-adding all available spectra of each star using a weighted least-squares regression with a cubic B-spline. This procedure achieves a precision at the level of 1 m/s \cite{serval}.
\begin{figure*}
    \centering
    \includegraphics[width=\textwidth]{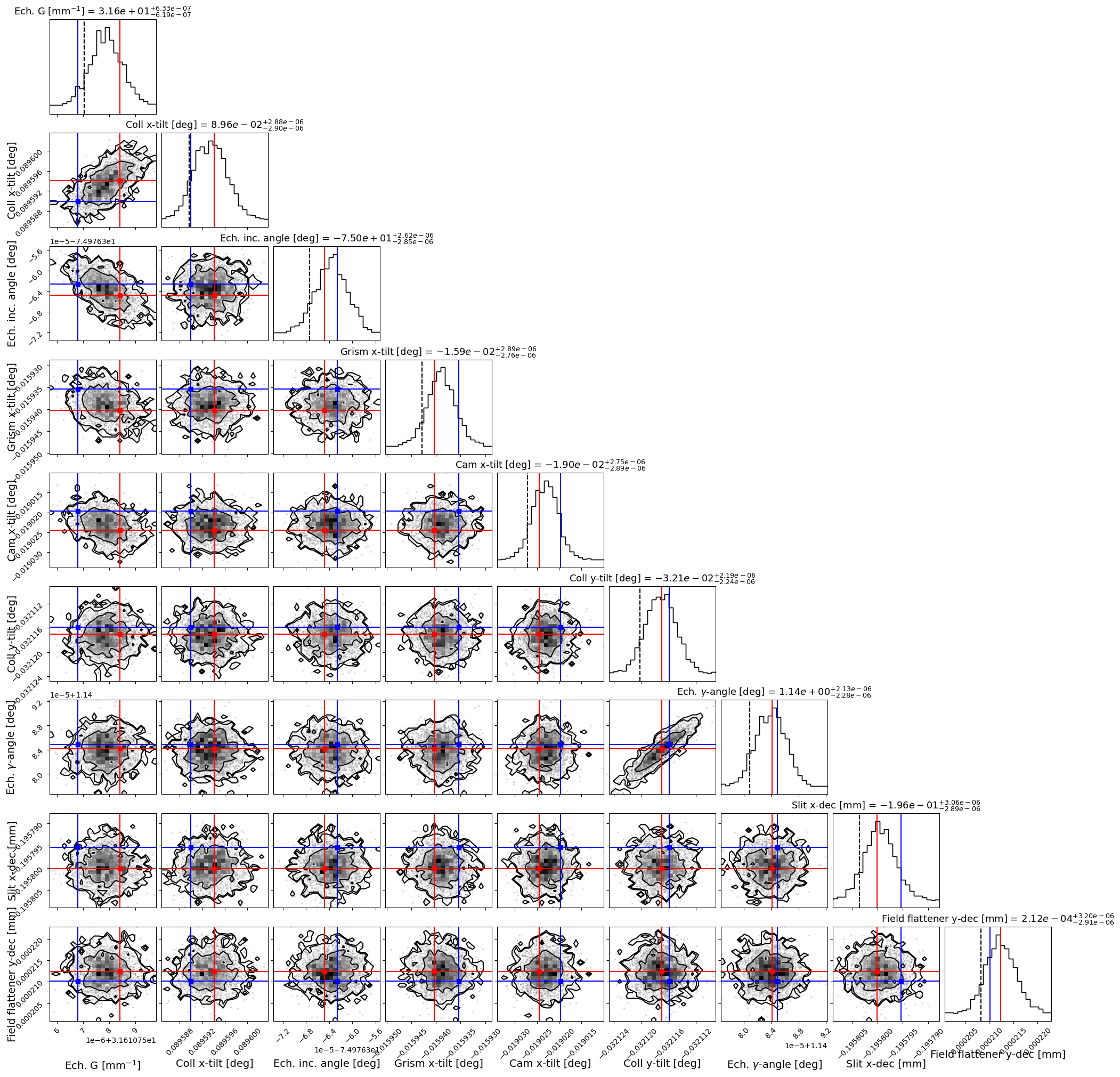}
    \caption{Corner plot of the most relevant parameters of the CARMENES VIS model constructed with {\tt moes}. The best fit parameters of the NS and SA optimizers are shown with blue and red lines, respectively.}
    \label{fig:cornerplot}
\end{figure*}

The instrument saw first light in late 2015, and more than 20\,000 spectra were obtained until the end of 2020 \cite{carmenes2020}. This data set resulted in several new planet detections, and it provides a long sequence of observations and daily calibrations. As our intent is to study how the optical path of light rays in the spectrograph affects the performance of the instrument, we will focus here on the spectrograph design. 

\subsection{Data description}\label{sec:data_description}
The CARMENES VIS channel uses three different HCLs: thorium-neon (Th-Ne), uranium-argon (U-Ar) and uranium-neon (U-Ne). We use the \cite{Redman2014} catalog for the Th lines, \cite{Sarmiento2014} for the U lines, and the NIST Atomic Spectra Database~\cite{Kramida2016} for the Ne and Ar lines. Our HCL spectral line list consists of a total of 15~481 spectral lines: 5~962 for U-Ne, 5~833 for U-Ar and 3~686 for Th-Ne. 

The {\tt caracal} pipeline provides the centroids $x_k$ and wavelengths $\lambda_k$ of the HCL lines on each of the spectral orders $o$ observed in the detector plane of the spectrograph.
{From the total line set we have selected a sub-sample of 8~589 HCL lines which are flagged by the \texttt{caracal} pipeline as good lines, e.g. lines that are not saturated and a have a reliable measured position in the detector plane (flag = 0, see table 4.5 in the \texttt{caracal} pipeline manual\footnote{\href{https://github.com/mtalapinto/moes/blob/main/carmenes/pipeline_manual.pdf}{\texttt{caracal} pipeline manual}}}). We use the spectral orders and the wavelengths of this sub-sample of lines as an input to our physical model, and we compare their measured positions on the detector with those modeled by \texttt{moes}.

Variations in the line intensities and shapes have a direct impact in the construction of the WLS. This effect is already considered and included in the calculation of the position error of each spectral line of the HCL spectra used in our analysis.

\subsection{Opto-mechanical sensitivity budget}\label{sec:error budget}

We have computed a sensitivity budget of the \texttt{moes} opto-mechanical parameter space with the goal of characterizing the effect of each parameter on the $(x, y)$ coordinates of the spectral lines. This information can then be used to optimize the computation time of the fitting procedure.

We have varied each parameter by an amount given by its thermal expansion (see Eq.~(\ref{eq:cte})) by considering a change in temperature of 0.3$^{\circ}$C. Distances between the optical elements are related to the thermal expansion of the aluminum optical bench, while the variations in the echelle grating groove spacing will be related to the expansion of Zerodur. In the case of the cross-dispersing grisms we consider the change in the groove spacing of its output surface as given by the thermal expansion of glass. Decentering and tilts of the different optical elements are given also by the thermal expansion of aluminum, as their mounts are made out of this material.

\begin{figure}
    \centering
    \includegraphics[width=0.85\textwidth]{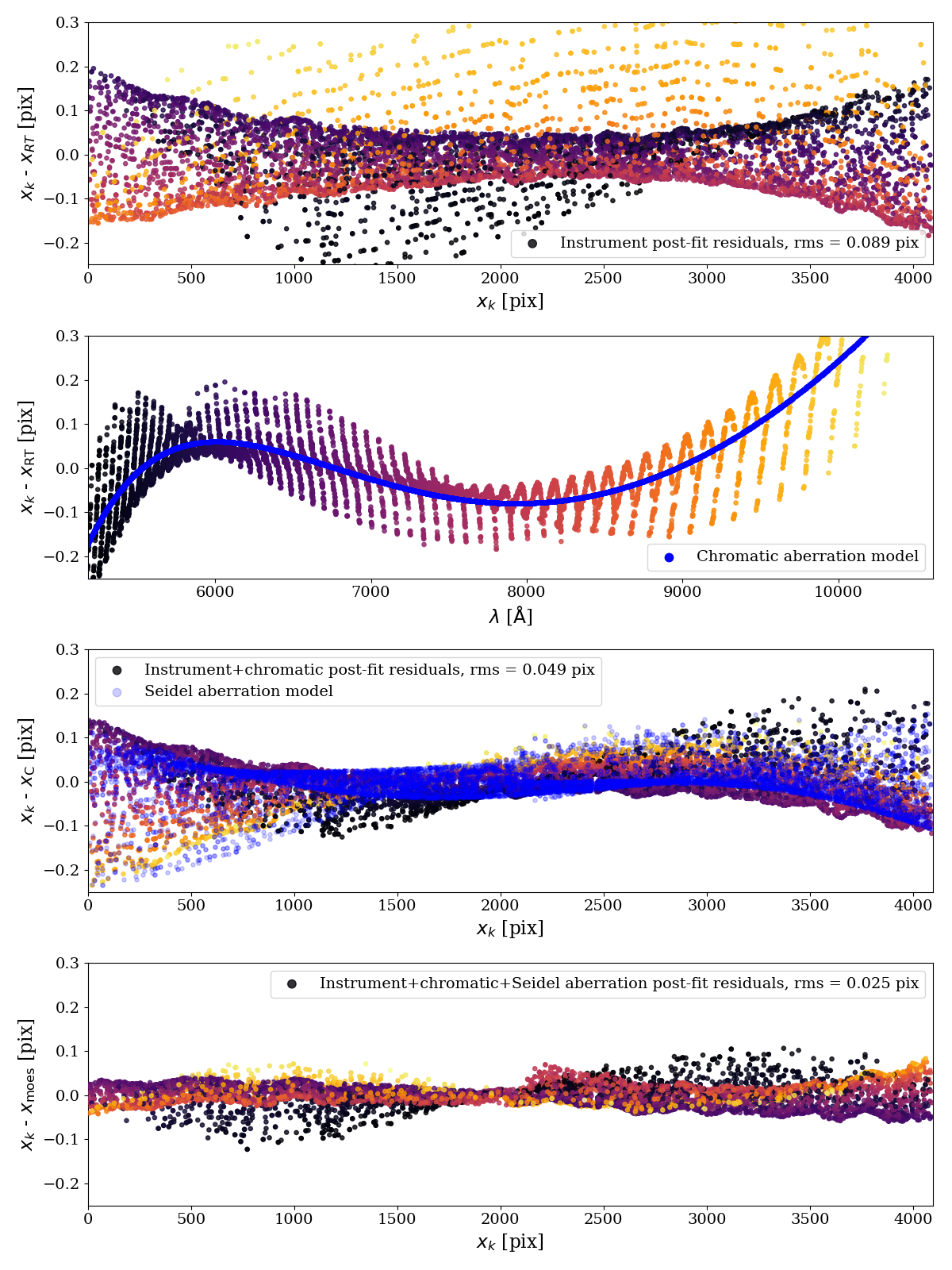}
    \caption{Residuals of the {\tt moes} fit of the VIS channel. Points are color-coded by wavelength according to Fig.~\ref{fig:echellogram}. From top to bottom the panels show: the residuals of the best-fit instrument model without aberrations as a function of the detector x-position, the residuals of the best-fit instrument model without aberrations as a function of wavelength, the best-fit instrument model corrected by chromatic aberrations as a function of detector x-position and the best-fit instrument model corrected by chromatic and Seidel aberrations as a function of the detector x-position. In the second panel from the top we have overplotted the chromatic aberration model in blue, while in the third panel we have overplotted the Seidel aberration model.}
    \label{fig:vis_residuals}
\end{figure}

Table~\ref{tab:error_budget} shows the nine most relevant instrumental parameters in the CARMENES VIS spectrograph, together with the amount $\delta p$, {computed using equation \ref{eq:cte}},  by which each parameter $p$ has been perturbed and its average effect on the $x$- and $y$-position of the spectral lines, $\delta x$ and $\delta y$, respectively. {With the exception of the echelle grating groove spacing $G$, the $\delta p$ for the parameters is given by the thermal expansion of aluminum 5083\footnote{\href{https://www.nist.gov/mml/acmd/aluminum-5083-o-uns-a95083}{Aluminum 5083 datasheet at NIST}}, which is the material used to build the mechanical mounts of the spectrograph. The $\delta p$ value for $G$ is given by the thermal expansion of Zerodur\footnote{\href{https://mss-p-009-delivery.stylelabs.cloud/api/public/content/3395a24104204531be2839378a783e3d?v=bca70f55&download=true}{Zerodur datasheet}}, the substrate of the echelle grating.}

The most relevant parameter is the echelle grating groove spacing $G$. {An expansion (or contraction) of the grooves of the echelle grating produces a change in the spectral line positions of about 4~pix along the echelle dispersion direction}. This is expected, as this is the main dispersion element. 
The next most relevant parameters are {the collimator x-tilt and echelle blaze angle. Both of them produce changes in the spectral lines of about half a pixel}.
The collimator x-tilt {relevance is mainly} driven by the white-pupil optical design of the spectrograph and the three reflections in the collimator. This makes the collimator a very sensitive optical element, as its orientation affects the direction in which the rays are reflected towards the echelle grating, the transfer mirror and the cross-dispersing grism. The echelle grating orientation has an impact on the spectral line positions of the same order of magnitude as the collimator, most likely due to its direct influence on the light dispersion.

We define $\delta x_T = x_{\rm moes}(T+\delta T) - x_{\rm moes}(T)$, i.e., $\delta x_T$ is the change in the $x$-position of a given spectral line after changing the temperature by $\delta T = \pm0.3~$K. Similarly, we define $\delta y_T = y_{\rm moes}(T+\delta T) - y_{\rm moes}(T)$  for the $y$-position. Figure~\ref{fig:error_bud_vis} shows the change in the $x$-positions as a function of wavelength for nine instrumental parameters: echelle grating groove spacing~$G$, collimator x-tilt, echelle blaze angle, grism x-tilt, camera x-tilt, collimator y-tilt, echelle~$\gamma$-angle, slit x-decenter {and field flattener y-decenter}. 
The same change in temperature affects multiple optical elements, and each one of them causes wavelength shifts with a specific amplitude and pattern. In some cases, several elements cause shifts that are similar but of very different magnitude, e.g., the {slit x-decenter} produces a similar effect to the {echelle grating constant $G$}, but {three} orders of magnitude weaker.

The model optimization is done in a sequence defined partially by the sensitivity that each parameter has on the position of the spectral lines. By inspecting the fit residuals during this sequential process and given the information provided by the sensitivity budget, we can decide which parameter to optimize next. Details of the optimization procedure will be described in Sect.~\ref{sec:fit}.


\subsection{Fitting model parameters to the calibration data}\label{sec:fit}

We fit the parameters of the physical \texttt{moes} model to the CARMENES VIS calibration spectra; this matches the instrument model to the instrument as built. The instrumental parameters considered cover various aspects of the spectrograph physics, e.g., spatial position and orientation of the manufactured optics, glass properties, and environment conditions. The optimization procedure adjusts the instrumental and aberration parameters in the model to match the calibration data obtained with the actual instrument. 

\begin{figure}
    \centering
    \includegraphics[width=0.85\textwidth]{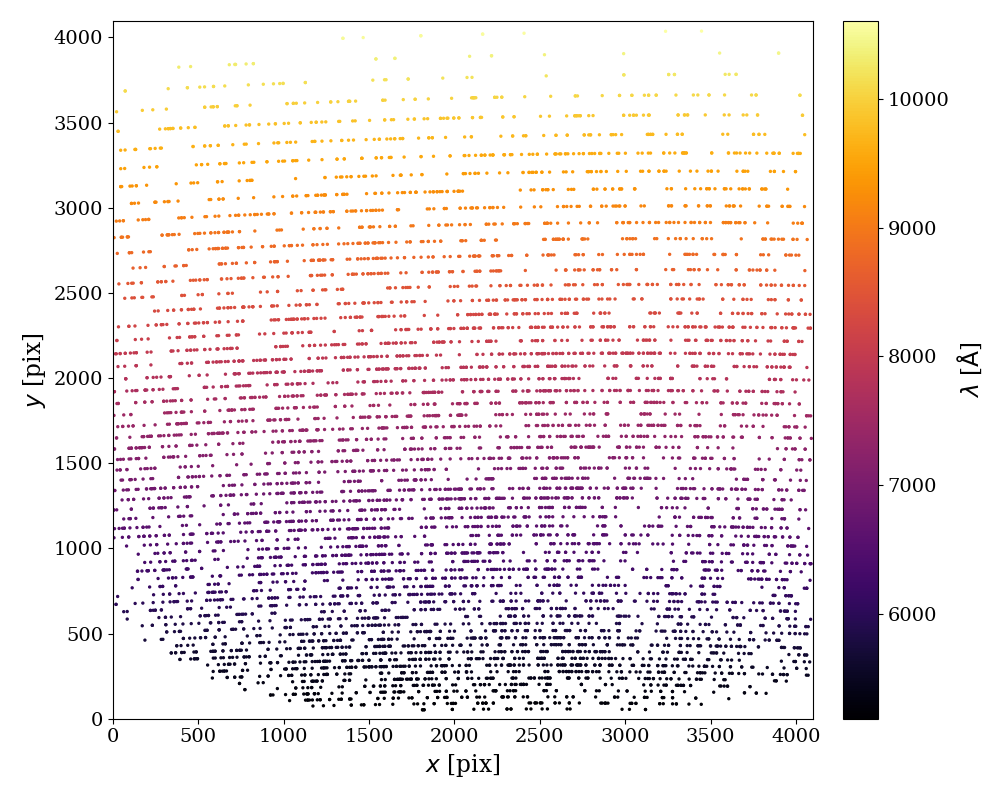}
    \caption{Echellogram of the CARMENES VIS spectrograph showing the position of the spectral lines in the detector plane modeled by \texttt{moes}. The modeled positions $(x, y)$ result from the instrument model including chromatic and optical aberrations. The wavelength is color-coded.
    }
    \label{fig:echellogram}
\end{figure}
The best-fit instrument parameters are obtained using a simulated annealing (SA) algorithm \cite{Kirkpatrick1983}. The main advantages of using SA are its ability to deal with highly non-linear models and to approach the global minimum without getting stuck in a local one. It is also quite versatile, since it does not rely on any restrictive properties of the model. Implementation of simulated annealing procedures requires choosing parameters such as the initial and final temperatures, the rate of \textit{cooling}, and number of function evaluations at each temperature. For this work we employ an SA routine\footnote{\url{https://github.com/perrygeo/simanneal}} that allows to adjust the initial and final temperatures, as well as the numbers of steps in each iteration. We set the initial temperature at $\tau_{\rm i}=25\,000$ and the final temperature at $\tau_{\rm f}=10^{-6}$, evaluating the function 5000 times at each temperature. We emphasize that, in this context, the temperatures are just parameters of the procedure and do not represent a physical entity. We use the default initial positions and directions for all the rays traced.

\begin{figure*}
    \centering
    \includegraphics[width=0.95\textwidth]{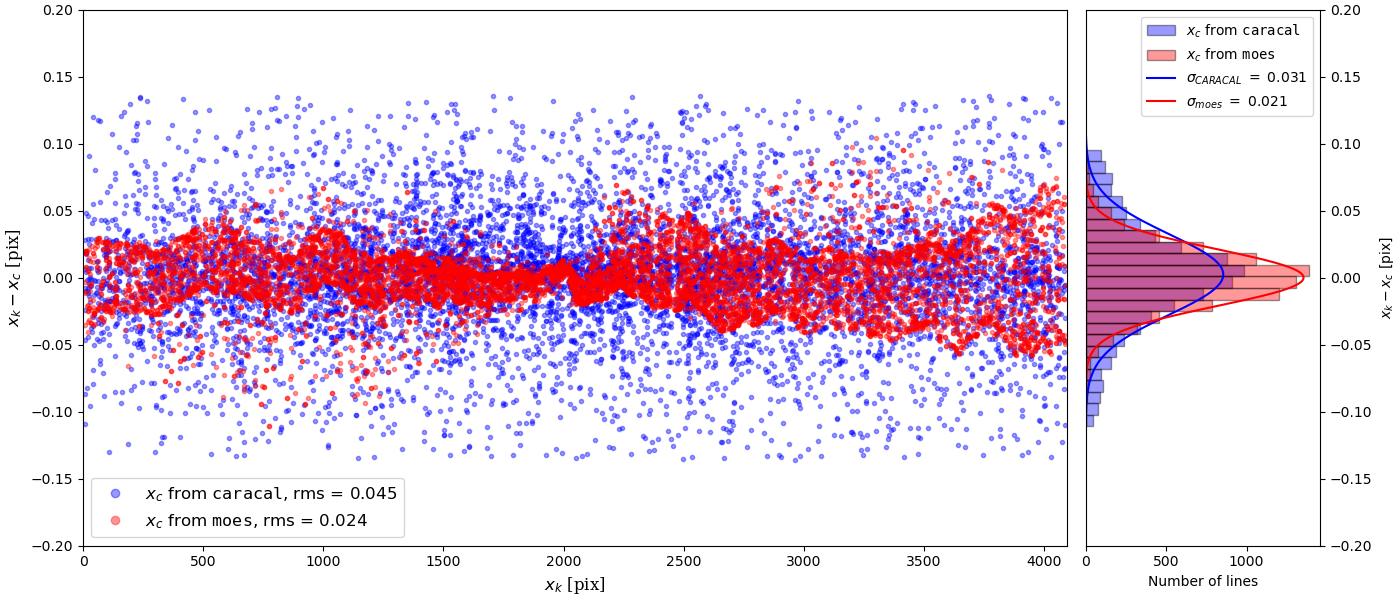}
    \caption{\textit{Left panel}: Comparison of the residuals from the wavelength solutions of {\tt caracal} (blue circles) and of \texttt{moes} (red circles) as a function of 
    wavelength. \textit{Right panel}: Distribution of the residuals for the \texttt{caracal} (blue) and \texttt{moes} WLS (red). We have overplotted the respective Gaussian fits in blue and red for the \texttt{caracal} and \texttt{moes} residuals. For both models we plot the same subset of lines defined in section \ref{sec:data_description}.}
    \label{fig:WLS_compare}
\end{figure*}
The SA probability function is
\begin{equation}
    _\mathcal{P} = \exp\left(-\frac{\chi}{\tau}\right)
\end{equation}
with 
\begin{equation}
    \chi = \sqrt{\frac{\chi_x^2 + \chi_y^2}{n_l}}~~,
\end{equation}
where $n_l$ is the number of spectral lines used in the fit, and $\chi_x$ and $\chi_y$ are given by
\begin{equation}\label{eqn:delta_x_siman}
    \chi^2_x = \sum_k \frac{[x_k - x(o_k, \lambda_k, \Vec{P}, \Vec{T})]^2}{\sigma_{x,k}^2} 
\end{equation}
and
\begin{equation}\label{eqn:delta_y_siman}
    \chi^2_y = \sum_k \frac{[y_k - y(o_k, \lambda_k, \Vec{P}, \Vec{T})]^2}{\sigma_{y,k}^2}~~,
\end{equation}
where $k$ is an index associated with each spectral line traced, $x_k$ and $y_k$ are the coordinates of the spectral lines measured on the detector in pixels, with uncertainties $\sigma_{x,k}$ and $\sigma_{y,k}$, respectively. $x(o_k, \lambda_k, \Vec{P}, \Vec{T})$ and $y(o_k, \lambda_k, \Vec{P}, \Vec{T})$ are the coordinates of the spectral lines calculated by \texttt{moes}, which are a function of the wavelength $\lambda$, the spectral order $o$ and a set of instrumental and environmental parameters $\Vec{P}$ and $\Vec{T}$, respectively. The values of $x_k$, $y_k$, $\lambda_k$, $o_k$ and $\Vec{T}$ are provided by the CARMENES reduction pipeline and are obtained during the daily calibrations.

We have set up an algorithm for the SA fitting routine that mimics the optical path of single rays through the spectrograph. Due to the complexity of the physical model, this process is done in stages. 
First we optimize each parameter individually. Then, we optimize the parameters sequentially from slit to detector, adding parameters to the optimization routine as the light rays pass through the different optical elements. Therefore, we start with the slit position and orientation, then we add the collimator x- and y-tilt, then we add the echelle grating parameters and so on, until we reach the detector. After this procedure, we observe the residuals to look for patterns similar to the ones observed in the position shifts for the different parameters during the computation of the sensitivity budget, in which case we optimize the parameters that could produce such pattern. The final step of the optimization procedure includes all 45 parameters of the spectrograph model.

To estimate the uncertainties of the parameters we investigate the posterior distribution of the instrumental parameter space using the nested sampling (NS) algorithm implemented in the \texttt{dynesty} package \cite{dynesty}. 
Nested sampling \cite{Skilling04, Skilling06} is a Monte Carlo algorithm aimed at efficient evaluation of the Bayesian evidence, providing posterior inferences as a by-product. The two main parameters of the NS algorithm are the \textit{priors} and the \textit{likelihood}, which define the boundaries in the parameter space and the probability of the model given the data, respectively. 
The definition of the priors is given by the nature of the different parameters. In our model we have identified four types of parameters $p$: tilts, decentering and positions of the optical elements, and the groove spacing of the dispersing elements -- the echelle grating and the back surface of the grism. 
In all cases we use uniform priors, so that the prior probability function is constant. This can be interpreted as all possible values being equally likely a priori. We assume this as we do not have any information regarding the true values of the parameters that produce a given dataset. Therefore, we assume constant priors with minimum and maximum values centered at the parameter value, extracted from the optical design of the spectrograph, in order to limit the range of the parameter space and optimize the time needed to explore it. 
\begin{figure*}
    \centering
    \includegraphics[width=\textwidth]{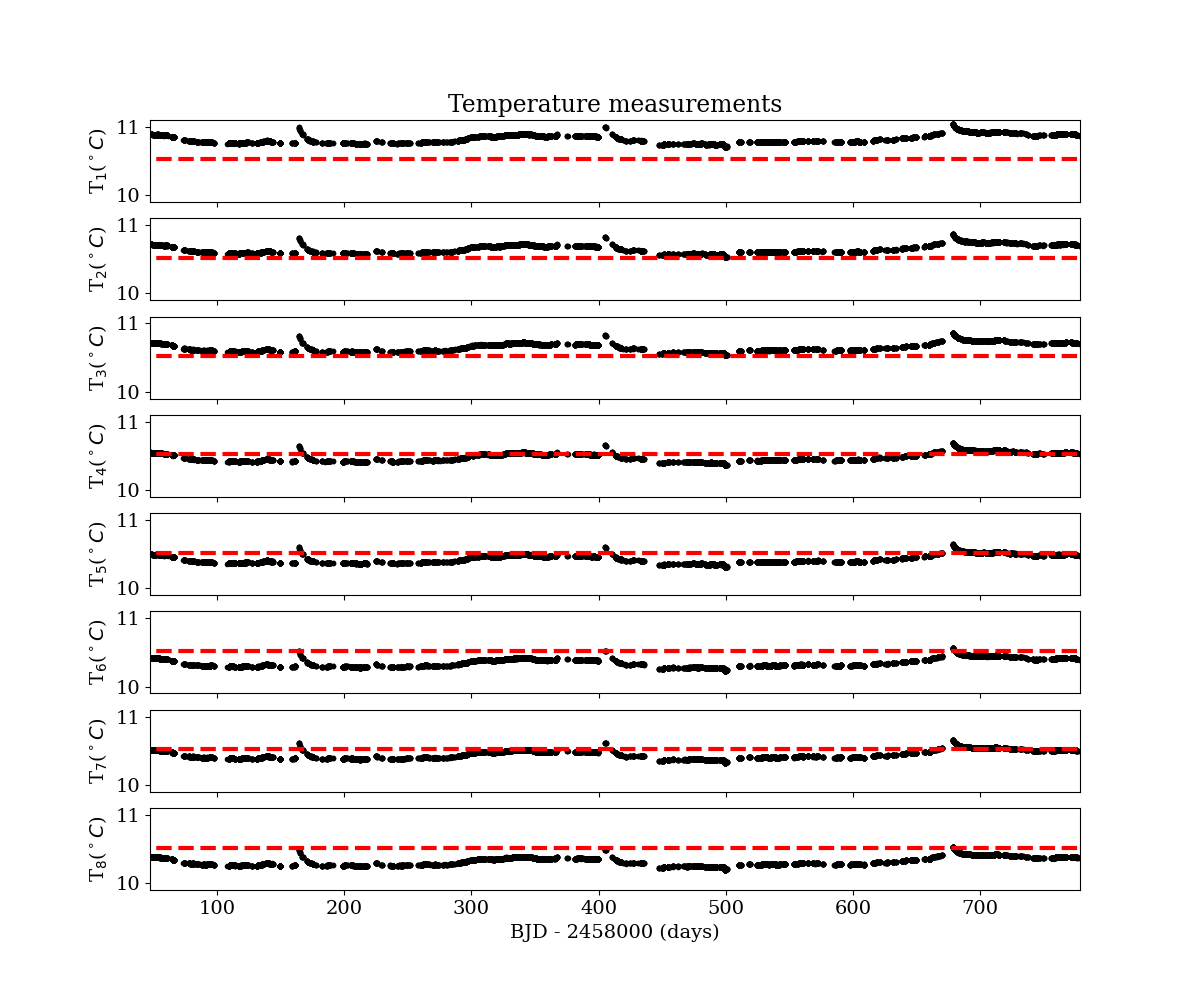}
    \caption{Time series of the temperatures measured by the sensors installed in the optical bench. We have marked the average temperature of the time series with a red line.}
    \label{fig:temperatures}
\end{figure*}

The likelihood is, for mathematical convenience, a log-likelihood defined by 
\begin{equation}
    -\ln\mathcal{L} = \chi_x^2 + \chi_y^2 + \ln 2\pi\sum_k (\sigma_{x, k}^2 + \sigma_{y, k}^2)~~.
\end{equation}
Figure~\ref{fig:cornerplot} shows the posterior distribution of the 9 most relevant instrumental parameters with the best fit parameters of the NS and SA solutions. One of the advantages of the NS algorithm is that by studying the posterior distribution one can identify possible correlations between the parameters.

Once we have found the best instrumental parameters, we calculate chromatic and optical aberrations corrections. 
As shown in Fig.~\ref{fig:vis_residuals}, we observe a strong dependence on wavelength of the instrumental model residuals. Hence, we first fit for chromatic aberrations and then for mono-chromatic aberrations. We emphasize that mono-chromatic aberrations are fitted to the residuals of the instrumental model including the chromatic aberration correction, and not fitted simultaneously with the chromatic aberration equations.

For the estimation of the aberration coefficients we also use \texttt{dynesty}. In the case of chromatic aberrations we fit the aberration function to the residuals of the instrument model, so
\begin{equation}
    x_k - x_{\rm RT}(o_k, \lambda_k, \Vec{P}, \Vec{T}) = \delta x_{\rm C}(\lambda_k)~~,
\end{equation}
where $x_{\rm RT}$ is the best-fit without aberration for $x$-position in pixels. The $x$-position after correcting for chromatic aberrations is given by
\begin{equation}
    x_{\rm C} = x_{\rm RT} + \delta x_C~~.
\end{equation}

To correct optical aberrations we fit the aberration function to the residuals of the instrument model including the chromatic aberration corrections, therefore
\begin{equation}
    x_k - x_{\rm C}(\lambda_k) = \delta x_{\rm O}.
\end{equation}

\begin{figure*}
    \centering
    \includegraphics[width=\textwidth]{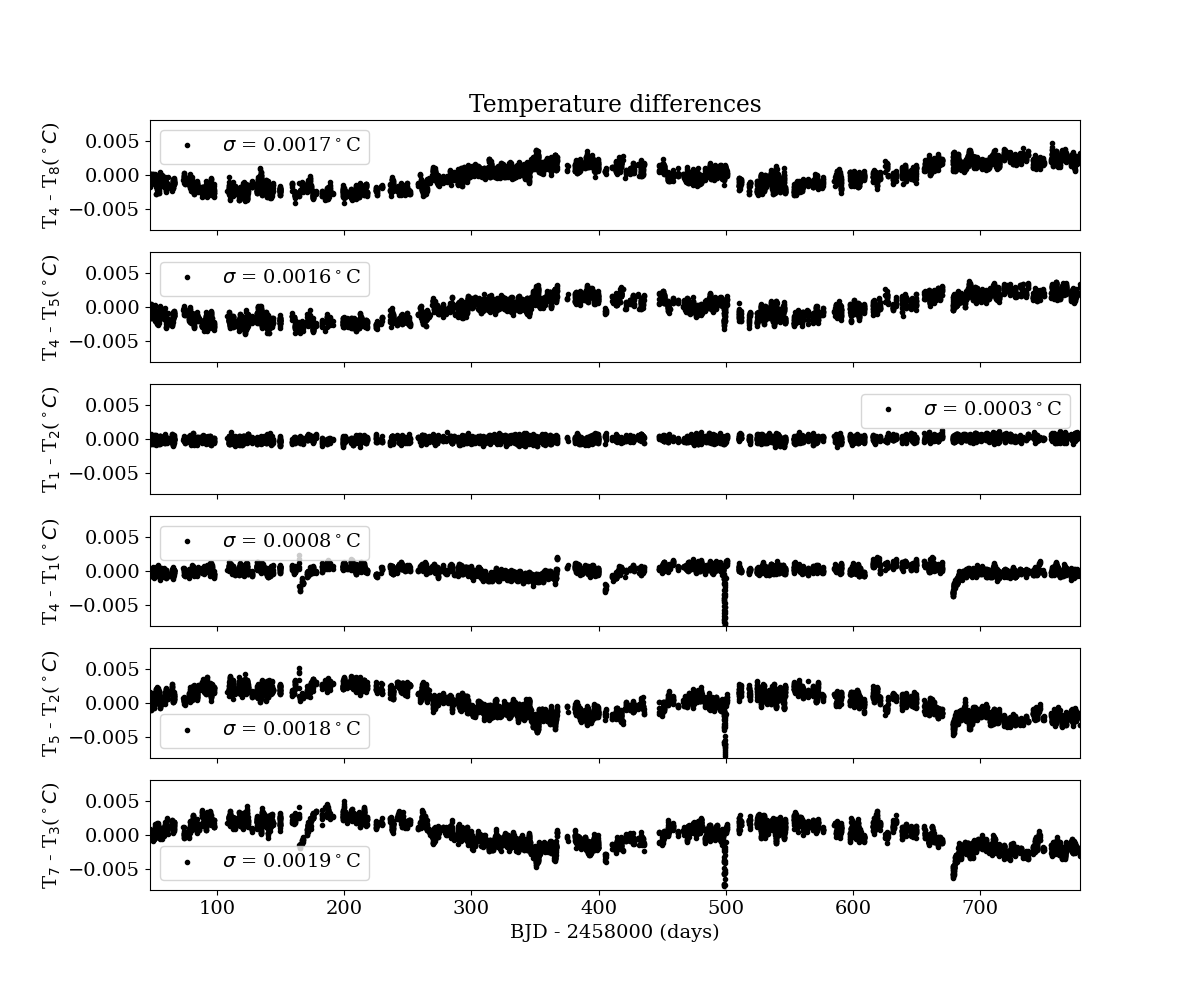}
    \caption{Time series of the temperature differences measured by the sensors installed in the optical bench. We also show the standard deviation $\sigma$ of each time series.}
    \label{fig:dtemperatures}
\end{figure*}

\subsection{Wavelength solution from ray tracing}

Figure~\ref{fig:vis_residuals} shows the fit residuals of the \texttt{moes} model for CARMENES VIS, obtained during the optimization process. 
The trend observed in the residuals of the instrument model as a function of wavelength (second panel from the top in Fig.~\ref{fig:vis_residuals}) strongly suggest the need of a chromatic aberration correction.
We observe significant improvements in the rms of the residuals as we include this in the model.  
After correcting for the chromatic aberrations, the rms is 0.049\,pix, and after correcting for optical aberrations the rms is reduced to 0.024\,pix. Figure~\ref{fig:echellogram} shows the $(x,y)$ positions of the HCL lines of the best fit model.

\begin{figure*}
    \centering
    \includegraphics[width=\textwidth]{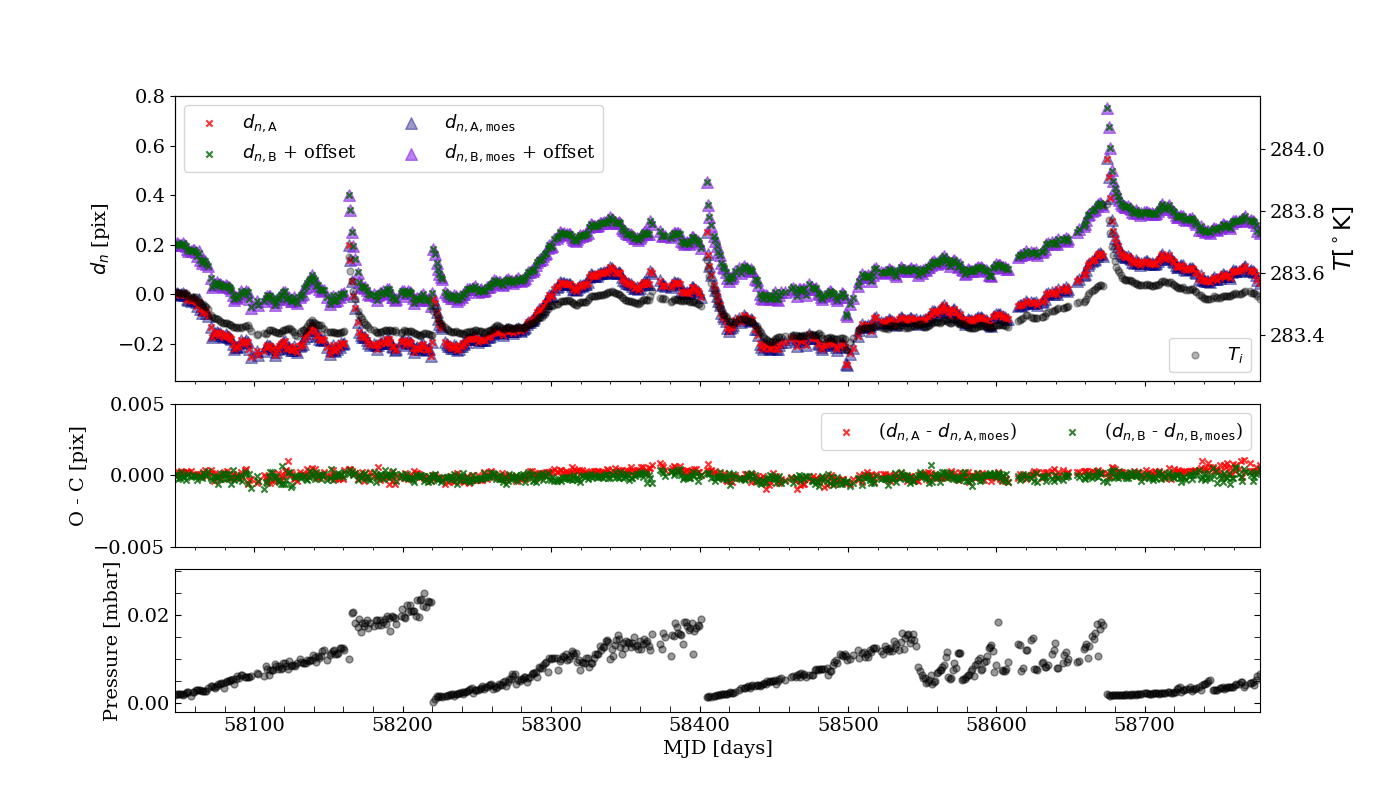}
    \caption{\textit{Top panel}: The drift values $d_n$ for fibers A (red crosses) and B (green crosses) measured using the position of the HCL spectral lines. The $\texttt{moes}$ model $d(t_n)$ for fibers A (blue triangles) and B (purple triangles) is overplotted. The fiber B drift are offset for visualization purposes. The right axis refers to the average temperature $T_n$ of the spectrograph (black circles). \textit{Middle panel}: The difference between the measured $d_n$ and predicted values of $d(t_n)$ for fiber A (red crosses) and B (blue crosses). \textit{Bottom panel}: Time series of the pressure measured inside the vacuum tank, in which the spectrograph is enclosed.}
    \label{fig:fiber_drifts}
\end{figure*}
          
Figure~\ref{fig:WLS_compare} shows the residuals of the WLS obtained empirically with {\tt caracal} and from the modeling with \texttt{moes}. The \texttt{moes} WLS provides much better results. According to our model, one pixel in the detector plane corresponds in average to 1~235\,m/s in RV, meaning that 1\,m/s corresponds to about 0.0008\,pix. The rms of the {\tt caracal} residuals is 0.038\,pix, while the \texttt{moes} residuals yield a rms of 0.024\,pix, which corresponds to about 46.9~m/s and 29.6~m/s, respectively. The right panel in Fig.~\ref{fig:WLS_compare} shows the histograms of the residuals together with Gaussian fits to the histograms. The Gaussian fit to the {\tt moes} residuals is considerably thinner than the one of {\tt caracal}, with FWHM of 0.049~pix and 0.072~pix, respectively.

\begin{figure*}
    \centering
    \includegraphics[width=0.95\textwidth]{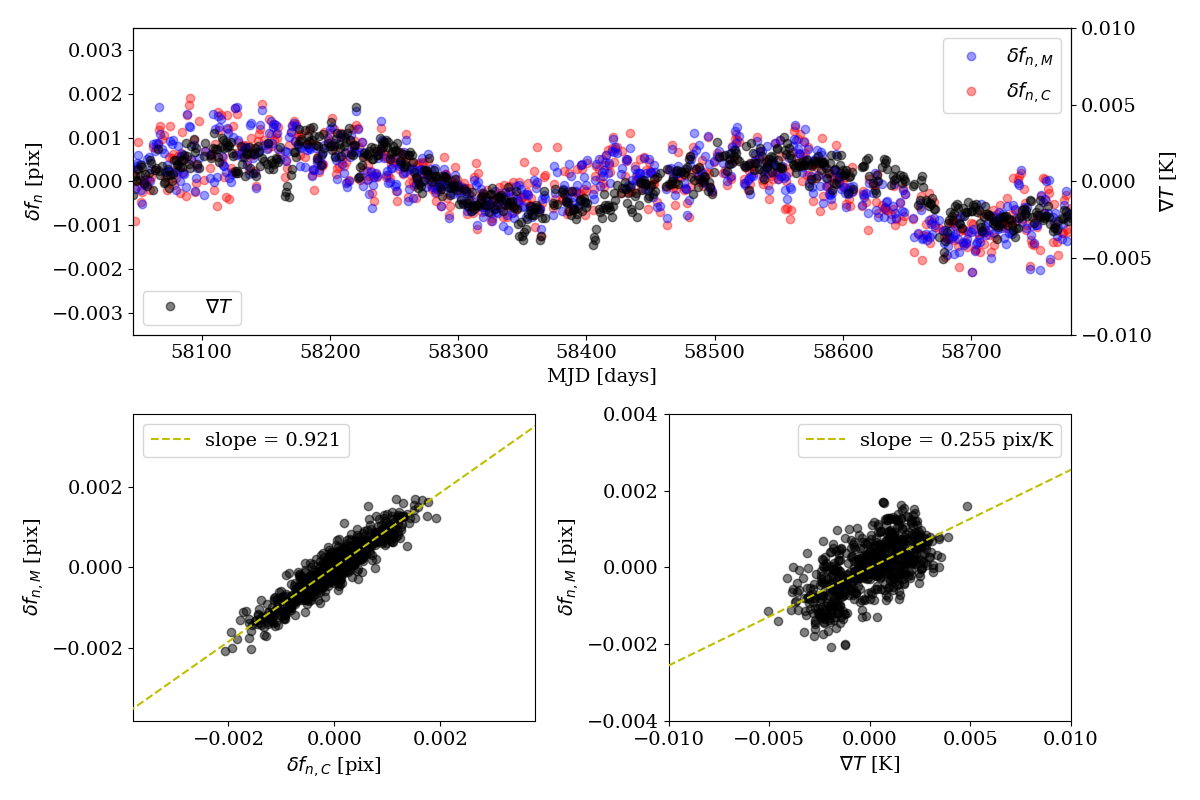}
    \caption{\textit{Top panel}: Time series of the differential fiber drifts $\delta f_{n}$ measured by {\tt caracal} (blue) and the prediction by \texttt{moes} (red). The right axis refers to the temperature gradient (black). \textit{Bottom}: Correlation plot of the  predicted differential drifts (\textit{left}) and temperature gradient (\textit{right}) with measured differential drifts.}
    \label{fig:ddplot}
\end{figure*}
The scatter of the residuals to the {\tt moes} model increases at the extremes of the wavelength range. This could be due to intrinsic limitations of the ability of the optical aberration polynomials to correct the aberrations near the edges of the focal plane.
While the actual value of the RV precision depends on many factors, the WLS rms serves a quantitative measure of the order of magnitude of the RVs that can be measured using our physical modeling approach. These results show that a ray tracing WLS can reach the accuracy required to measure precise stellar RVs. Moreover, our physical-model based WLS performs considerably better than the traditional polynomial WLS.

\section{Temperature-based Zero Points}\label{sec:TZP}

Each of the CARMENES spectrographs is fed with two octagonal fibers, placed next to each other, which carry the light of the observed star and the calibration sources -- HCLs and FPs etalons -- to the entry of the spectrograph. The image of both fibers is re-imaged by the FN-system in the image slicer to create a pseudo-slit before passing through the spectrograph optics. We will refer to fiber~A as the fiber that carries the stellar light and to fiber~B as the fiber that carries the calibration light.

As shown in Sect.~\ref{sec:error budget}, small changes in the instrument environment will induce changes in the opto-mechanics, affecting the position of the spectral lines on the detector. The CARMENES VIS spectrograph is hosted in a vacuum tank to minimize the effects of pressure, and it is passively thermally stabilized \cite{carmenes_vis_walter}. The temperature in the spectrograph is measured using eight contact temperature sensors located in different positions on the optical bench, as shown in Fig.~\ref{fig:spec_trace}. This allows to track the temporal evolution of the mean temperature of the instrument, but also to keep track of temperature gradients inside the spectrograph. 
To accurately model the environmental conditions of the spectrograph, the temperature of the different opto-mechanical components is assumed to be the one measured by their respective closest temperature sensor, e.g., the temperature of the collimator is taken to be the one measured by the sensors ${\tt TEMP\_}1$ and ${\tt TEMP\_}2$, while the temperature of the echelle is measured by ${\tt TEMP\_}8$. With this approach we can provide an accurate description of the instrument and, more importantly, we can implicitly include the effects of temperature gradients inside the spectrograph in the WLS calculation. 
Figure~\ref{fig:temperatures} show the time series of the temperatures measured by the sensors installed in the optical bench. While the temperatures measured in the different sensors follow the same trend, not all of them measure the same temperature, which may be indicative of temperature gradients in the spectrograph.
Therefore, we computed the temperature differences between some of the sensors to study such gradients, which are shown Fig.~\ref{fig:dtemperatures}. The top three plots show the top-differences between top-to-bottom located sensors, while the bottom three plots show the right-to-left temperature differences measured at the different sensors in the optical bench.

\subsection{Fiber drifts}

To study the effect of temperature and pressure on the spectral line positions we have computed the \textit{fiber drifts}, defined as
\begin{equation}
    d_{n} = \frac{1}{N_n}\sum_{k=1}^{N_n} d_{n, k}
\end{equation}
with
\begin{equation}
    d_{n, k} = x_{n, k} - x_{0,k}~~,
\end{equation}
where $N_n$ is the total number of spectral lines for a given date $n$, $d_{n, k}$ is the drift of a single line $k$ in pixels at a given time $t_n$, $x_{n, k}$ is the measured position of spectral line $k$ at time $t_n$ and $x_{0, k}$ is the measured position of spectral line $k$ at time $t_0$. The times $t_n$ and $t_0$ correspond to the times at which the WLS for the observing night $n$ and the first observing night ($n{=}0$) considered in our sub-sample was created.

The instrument has been operational since January 2016; however, we have selected a sub-sample of dates between the 20th of October 2017 (JD$=$2\,458\,046) and the 20th of October 2019 (JD$=$2\,458\,777), covering two full years of observations.


The top panel in Fig.~\ref{fig:fiber_drifts} shows the fiber drift for fibers A and B measured from the position of the HCL spectral lines used in the WLS creation and modeled using \texttt{moes}. We have overplotted the average of the temperature measured by the eight sensors on the spectrograph bench for comparison.

The temperature fluctuates around $\sim 283.3 ^{\circ}$K with several jumps. These jumps are produced by the regeneration of the sorption pumps of the detector cryostat and of the spectrograph vacuum tank. The goal of this intervention is to restore the vacuum conditions inside the detector cryostat and the vacuum tank. Every time the sorption pumps are regenerated, the temperature rises almost instantly by a few 0.1$^{\circ}$C and then slowly decreases, reaching equilibrium values after a few days.
Although we can clearly observe the effect of these interventions in the pressure time series, shown in the bottom panel of Fig.~\ref{fig:fiber_drifts}, we note that this is the pressure measured inside the spectrograph, which is not the same as the detector cryostat pressure. 
These jumps in temperature and pressure have a direct impact on the position of the spectral lines on the detector. As shown in Fig.~\ref{fig:fiber_drifts}, the drift time series $d_n$ follows the same trend as the average temperature $T_n$, while the jumps in the $d_n$ values coincide with those of pressure and temperature. The middle panel in Fig.~\ref{fig:fiber_drifts} shows the residuals of the fiber drifts. These residuals have been calculated by subtracting the measured values of $d_n$ from the ones modeled by \texttt{moes} $d(T_n)$. The rms of the fiber A residuals is 5.9$\cdot$10$^{-4}$\,pix and the rms of the fiber B residuals is 3.9$\cdot$10$^{-4}$\,pix.

We emphasize that the modeled fiber drifts are calculated from a set of physical parameters that do not change considerably from one day to the next, based on the assumption of the inner mechanical stability of the instrument. Therefore, our model is valid for thermal equilibrium conditions. Drastic changes would require the modeling of thermal inertia.

Even so, these results show that our ray tracing wavelength solution can actually provide a physical model that accurately  describes the observed drifts of the spectral lines in the detector plane of the spectrograph.

\subsection{Differential fiber drift}\label{sec:diff_drift}


To characterize the evolution of the wavelength positions of the science and calibration spectra, we have defined the differential fiber drift as
\begin{equation}
    \delta f_n = \frac{1}{N_n}\sum_{k=1}^{N_n} \delta f_{n, k}
\end{equation}
with
\begin{equation}
    \delta f_{n, k} = (x_{{\rm A},k} - x_{{\rm B},k})_n~~,
\end{equation}
where $x_{{\rm A},k}$ and $x_{{\rm B},k}$ are the wavelength positions for the science and calibration spectra in pixels, respectively, $\delta f_{n, k}$ denotes the difference between the x-positions of the science and calibration spectra, and $\delta f_{n}$ is the average of $\delta f_{n, k}$ for a given night $n$.
To better understand the nature of the observed $\delta f$, we have calculated the temperature gradient within the spectrograph, defined as 
\begin{equation}\label{eq:dt}
    \nabla T = \frac{T_1 + T_2 + T_3}{3} - \frac{T_6 + T_7 + T_8}{3}~~,
\end{equation}
where $\nabla T$ is the temperature gradient, and $T_j$ are the temperatures measured by sensors ${\tt TEMP}\_j$ at the positions indicated in Fig.~\ref{fig:spec_trace}. $T_1$, $T_2$, $T_3$ are the sensors on the collimator side, while $T_6$, $T_7$ and $T_8$ are the sensors on the detector side. We defined the temperature gradient based on the results of the sensors temperature differences from Fig.~\ref{fig:dtemperatures}, that show that the highest amplitude of temperature differences seems to come from a side-to-side gradient, than to a top-to-bottom gradient.

Figure~\ref{fig:ddplot} shows the differential drift time series measured by {\tt caracal}, $\delta f_{n, M}$, and computed with {\tt moes}, $\delta f_{n, C}$. For each time series, each point corresponds to an observing night $n$, with the exact time corresponding to the time the WLS of the $n$-night was created. Therefore, the temperatures and pressure considered when calculating $\delta f_{n, C}$ are the ones of the instrument at these times. We have centered the $\delta f_n$ on the mean, with $\overline{\delta f}_{n, {\rm M}}$
=~0.077~pix and $\overline{\delta f}_{n, {\rm C}}$~=~0.078~pix. The temperature gradient has also been plotted for comparison.
\begin{figure*}
    \centering
    \includegraphics[width=0.95\textwidth]{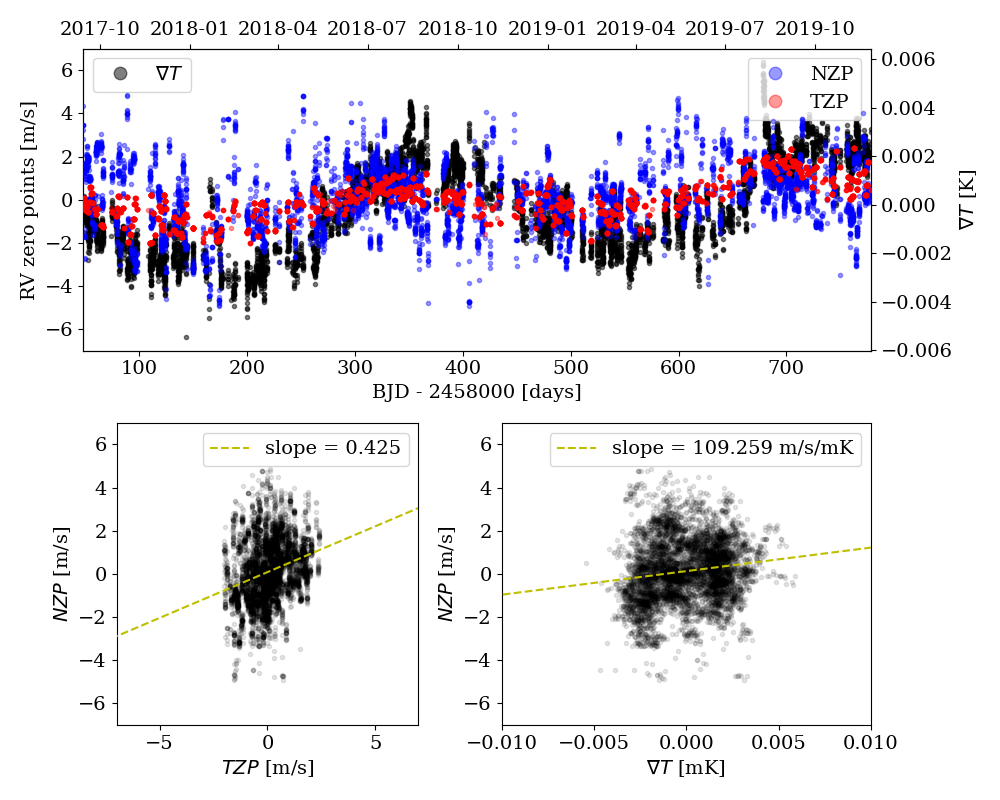}
    \caption{\textit{Top panel}: Time series of the TZP (red), NZP (blue), and $\nabla T$ (black, right axis). \textit{Bottom-left panel}: TZP versus the NZP. The best-linear fit, plotted in yellow, yields a slope of 0.489. \textit{Bottom-right panel}: $\nabla T$ versus NZP. The best linear fit yields a slope of 111.9 m/s/K.}
    \label{fig:tzpnzp}
\end{figure*}
Similar to what we observed in the time series of the fiber drifts, the \texttt{moes} predictions for the $\delta f_n$ are in excellent agreement with the $\delta f_n$ measured with {\tt caracal}. The slope of the best linear fit is 0.921, as shown in the bottom left panel in Fig.~\ref{fig:ddplot}. The rms of the $\delta f_{n, {\rm M}}$ values is 7.75$\times$10$^{-2}$~pix and the rms of the $\delta f_{n, {\tt C}}$ is 7.72$\times$10$^{-2}$~pix. If we correct the data for the model, the difference $\delta f_{n, {\rm M}} - \delta f_{n, \tt C}$ yields an rms of 7.1$\times$10$^{-4}$~pix. The Pearson coefficient of 0.93 indicates a strong positive correlation.

Moreover, the $\delta f_n$ values seem to follow a similar trend as the temperature gradient. The bottom right panel in Fig.~\ref{fig:ddplot} shows $\nabla T$ versus the $\delta f_{n, {\rm M}}$ values. The best linear fit yields a slope of 0.255~pix/K. The rms of the temperature gradients is 0.002$^{\circ}$K.
The Pearson correlation coefficient is 0.62, suggesting a moderate positive correlation. However, the strong correlation between the measured and modeled differential fiber drifts, together with the rms reduction after subtracting the model from the data, suggest that temperature gradients are the main cause for the observed differential fiber drifts.


\subsection{Ray tracing RV systematics}
 
Since the start of the observations, a number of instrumental effects and calibration issues affecting the data on the m/s level have been identified. Therefore, taking advantage of the survey-mode observations, for each night of guaranteed time observations (GTO), an instrumental nightly zero-point (NZP) of the RVs is calculated by using all the stars with small RV variability (RV-quiet stars) observed in that night. All RV measurements are corrected for this NZP.

The sample of RV-quiet stars was defined as the sub-sample of CARMENES-GTO stars with RV standard deviation $<10$\,m/s. The sample of RV-quiet stars includes $\sim$200 stars of which 10–20 are observed in a typical night \cite{nzppaper}.

The fact that we observe RV systematics in the sample of RV quiet stars, suggest that there might be instrumental effects that have not been taken into account during the RV computation. 

As mentioned in Sect.~\ref{sec:diff_drift}, whilst we can track the change in the $\delta f_n$ every day using the calibration data, it is not possible to track differential fiber drifts during the observing night, as no calibration data are obtained for the science fiber spectra during the night. To overcome this, we propose to use \texttt{moes} to predict how much the $\delta f_n$ has changed due to the changes in temperature and pressure.
In this context, we have defined the \textit{temperature-based zero point} (TZP) as
\begin{equation}\label{eq:tzp}
    {\rm TZP}_i = s\cdot[\delta f_{n,{\rm M}} + \delta f_{i, {\tt C}}(\Vec{T_i}) - \delta f_{n, {\tt C}}(\Vec{T_c})]~~,
\end{equation}
where $s$ is the pixel-to-rv scaling factor, and $\Vec{T_i}$ and $\Vec{T_c}$ correspond to the environmental vectors with temperatures and pressure measured at the observation and calibration times, respectively. The $i$-index refers to each single science observation.
We have noticed during our calculations that the variations of the pixel-to-rv factor with temperature, and consequentially with time, are well below the 1~m/s level. Therefore, we consider that there is no temperature nor time dependency in $s$.

The first term inside the parentheses considers the measured $\delta f_n$ at time $t_c$ using the calibration data, and the difference between the second and third terms corresponds to the change in the $\delta f_{n}$ produced by changes in temperature and pressure since time $t_c$ until the time the $i$th-observation occurs, as predicted by {\tt moes}.

For each night we have computed the \texttt{moes} WLS, setting the instrument environmental conditions to the ones measured at the time the CARMENES WLS is created, during the afternoon calibrations. Using the same set of instrumental and aberration parameters of the \texttt{moes} WLS, we calculate the pixel-to-rv scaling factor $s$ and the $\delta f_i$ as a function of the temperature and pressure measured at the time the $i$th-observation occurred that $n$-night.

We have considered each observation carried out between the 20th of October of 2017 and the 20th of October of 2019. The top panel in Fig.~\ref{fig:tzpnzp} shows the time series of the TZP plotted against the NZP and the temperature gradient $\nabla T$.
Similar to what we observe in the differential drifts, the NZP, the TZP and $\nabla T$ seem to follow the same trend. This result suggests that the NZPs may be driven, at least partially, by the temperature gradient fluctuations in the spectrograph.


While both type of RV systematics -- the NZPs and the TZPs -- follow the same trend, the NZPs show a larger scatter. The rms of the NZP time series is 1.64~m/s, while the rms of the TZP time series is 0.99~m/s. 
It is important to acknowledge that our modeling approach only receives the environmental conditions and the spectral information of the HCLs to model the TZPs. Moreover, the spectral information of the HCLs does not change drastically over the period of time we have considered in our investigation. Therefore, the TZPs are driven mainly by the changes in the spectrograph temperature and pressure. 

The NZPs are calculated using the stellar RVs measured each night, so the larger scatter could be explained by the instrument accuracy, including all sources of systematics, plus an intrinsic stellar component, which one could label as \textit{stellar jitter}. 

The bottom left panel in Fig.~\ref{fig:tzpnzp} shows the TZP versus the NZP. The best linear fit to these data yields a slope of 0.489. The Pearson correlation coefficient is 0.30. While this value suggests a weak correlation, it provides evidence supporting the link between both quantities, one of which comes from a purely physical model of the instrument.

The bottom right panel in Fig.~\ref{fig:tzpnzp} show the NZP versus $\nabla T$. The best linear fit to these data yields a slope of 111.9~m/s/K. The non-zero slope suggest that both quantities may be correlated. The Pearson coefficient value of 0.15 suggest a weak correlation. {While} the non-zero value of the slope provides evidence supporting the hypothesis that the observed NZP may be driven by the changes in the spectrograph temperature gradients, {this correlation is not significant enough to provide a robust claim}.

\section{Discussion}\label{sec:discussion}

\subsection{Instrument modeling}


The complexity of ray tracing models of echelle spectrographs induces an additional challenge when fitting the model to real calibration data. This is mainly due to the sequential nature of ray tracing, which entangles the different optical parameters. To overcome this limitation, it was proposed the use of an SA routine to fit the physical model of the X-shooter spectrograph to the calibration data\cite{Bristow06}. As discussed in Sect.~\ref{sec:fit}, SA routines provide an advantage when optimizing complex optical systems, as they allow to efficiently optimize highly non-linear models. The combination of the physical model of X-shooter with a high efficiency fitting routine allowed to establish a correlation between the instrumental parameters of the best-fit model with the flexures produced by the weight of some of the opto-mechanical components of the instrument, showing the strong predictive capabilities of physical models.

Here we also employ an SA routine to estimate the instrumental parameters that best fit the calibration data, and we use an NS algorithm to build the posterior distribution of the parameter space. The combination of both techniques allows to determine the best-fit parameters with their uncertainties, while at the same time evaluating possible correlations between the model parameters.

As seen in Fig.~\ref{fig:cornerplot}, in the CARMENES VIS spectrograph model we observe a strong positive correlation between the values determined for the y-tilt of the collimator and the echelle $\gamma$-angle. 
Furthermore, we observe a moderate positive correlation between the grating constant and the collimator x-tilt, and between the grating $\gamma$-angle and the collimator y-tilt. These correlations could be explained by the fact that the collimator and the echelle grating are sequential elements, and their tilts are along the same axis, but also because the beam is reflected in the collimator three times: after the FN-system, after the echelle dispersion and after the transfer mirror. Given the optical design of the instrument, which employs one single monolithic parabolic mirror as collimator, one could expect such correlations in the collimator's orientation parameters.

This is also observed in the spectrograph sensitivity budget, presented in Sect.~\ref{sec:error budget}, where we have been able to evaluate which instrumental parameters are most sensitive to temperature changes. With these data it is possible to constrain the parameter space explored by the model fitting routine, optimizing the model computation time. But more importantly, these results show that, in the CARMENES VIS spectrograph, the mechanical stability of the CCD and collimator orientation are essential for ensuring the spectral stability required to measure precision RVs.

Most physical models built for astronomical instruments have considered the inclusion of aberration terms in the ray tracing  modeling. Ballester et al. (1997)\cite{Ballester97} proposed the use of a low-order polynomial, reaching a precision of about 1~pixel. 
Bristow et al. (2006)\cite{Bristow06} employed the same algorithm as Ballester et al. (2007)\cite{Ballester97} for the physical model of X-shooter, while the approach of Chanumolu et al. (2015)\cite{chan15} consisted in a paraxial tracing component coupled with a set of Buchdahl aberration polynomials~\cite{buchdahl54, Buchdahl56, Buchdahl58} to model optical aberrations, matching the model to the data with an accuracy of about 0.08~pixel.
Our proposal consists in the use of a direct ray tracing algorithm with classical Seidel aberrations calculated directly at the detector focal plane~\cite{mahajan14}, matching the instrument model to the calibration data with an rms of about 0.021~pixel. 
These results show the importance of the inclusion of aberration functions in the computation of the ray tracing model, as evidence suggests that paraxial or direct ray tracing models alone are not capable of describing optical systems at the sub-pixel level.

Current state-of-the-art calibration pipelines provide wavelength solutions with different accuracies. The \texttt{ceres}\cite{ceres} pipeline yields an average accuracy in the FEROS spectrograph of about 0.06~pix, whereas ESPRESSO's WLS has shown wavelength-dependent discrepancies of about 24~m/s, which corresponds to about 0.038~pix in the high resolution mode (R$\sim$190,000)\cite{espresso_wls}. \texttt{moes} WLS residuals rms is 0.024~pix, setting a new standard to the precision of the WLS.
\begin{figure}
    \centering
    \includegraphics[width=\linewidth]{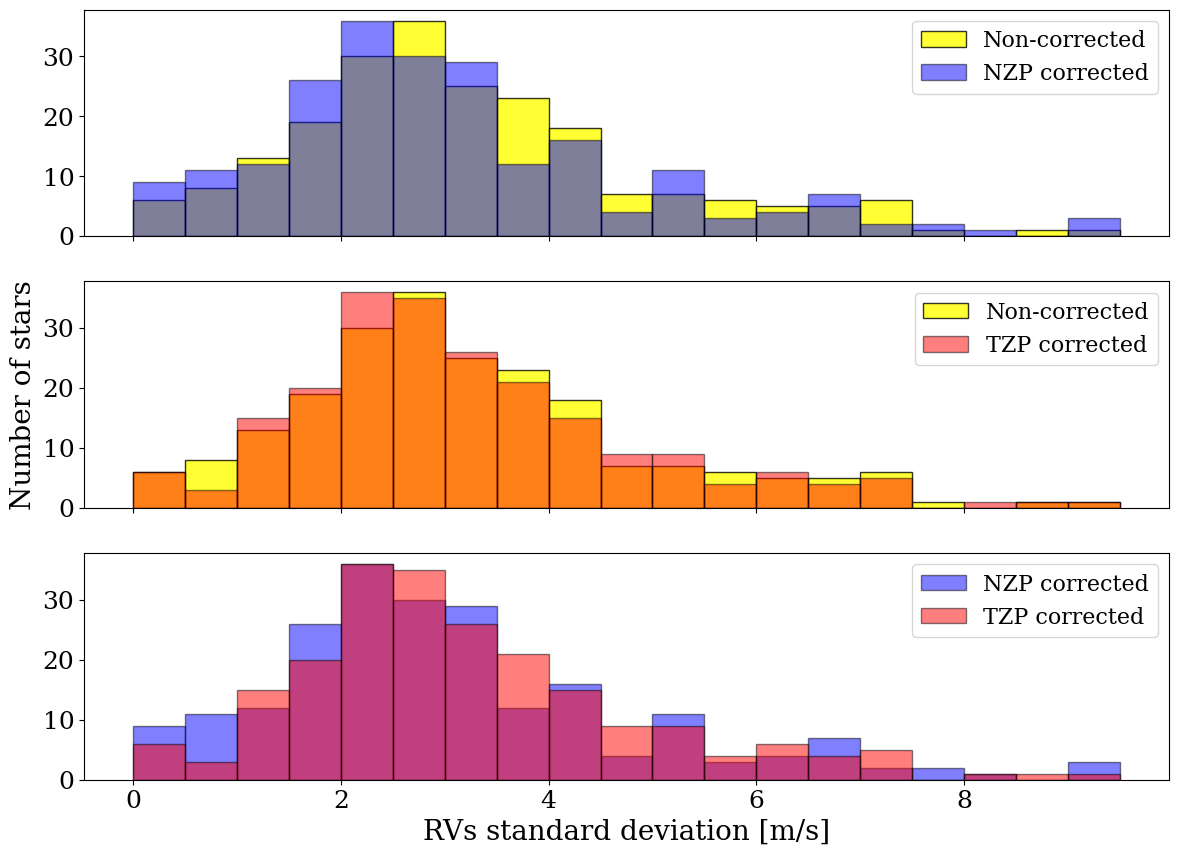}
    \caption{Histogram of the standard deviation of the CARMENES VIS RVs for a sample of RV-quiet stars. We show non-corrected stars in yellow, NZP-corrected stars in blue and TZP-corrected stars in red.}
    \label{fig:stdrvc}
\end{figure}

\subsection{TZP vs NZP}


The CARMENES VIS spectrograph is mounted on an optical bench inside a vacuum tank. The instrument operates at room temperature, with a variation amplitude of about 0.3$^{\circ}$C during the two years considered in our study (see Fig.~\ref{fig:fiber_drifts}). Given the conditions in which the instrument is housed, the only parameters that change with time are temperature and pressure. Based on this hypothesis, one can assume that the observed change in the positions of the spectral lines of the science and calibration fiber are driven mainly by changes in these two parameters. We must acknowledge, though, that the instrument may still be subject of external sources of noise, e.g. instrument interventions and/or misplacement of the science and calibration fibers at the front-end. Here we focus only in the systematics that may arise due to temperature and pressure fluctuations inside the spectrograph vacuum tank.

Using our ray tracing algorithm {\tt moes}, we calculate the change in the ($x$, $y$) positions of a set of wavelengths in the science and calibration spectra, as a function of the environmental parameters. We calculate these positions at the time (and temperature and pressure) the wavelength solution of the observing night is created, and at the time a given observation is conducted, to track the short-term changes $\delta f_n$. 

In this context we define the TZP (see Eq.~\ref{eq:tzp}) as the expected RV drift produced by the temperature and pressure changes in the spectrograph.
The comparison of the {\tt moes} TZP with the observed NZP, and with the measured temperature gradient in the spectrograph, shown in Fig.~\ref{fig:tzpnzp}, has provided quantitative evidence for the correlation between the observed NZP and the environmental changes of the spectrograph. Moreover, we can provide an estimate of the expected RV systematics as a function of temperature.

During the two-year period considered in our study, we have taken observations during 434 nights, yielding a total of 5422 RV measurements of 221 stars. To study whether the TZP values can be employed to improve the overall performance of the instrument, we corrected the measured RVs by the TZPs. Using {\tt moes}, we create a WLS for each observing night, we compute the TZP for each stellar observation, and we subtract it from the measured stellar RV. 

We consider that the TZP correction is successful if the standard deviation of the RV time series for a given target decreases after applying the correction. Of the 221 stars in our sample, 118 (53.4~\%) have standard deviations smaller than the ones of the non-corrected RV time series, and 23 (10.4\%) stars show standard deviations smaller than the ones of the NZP-corrected RV time series. Figure~\ref{fig:stdrvc} shows the distribution of the standard deviation for the non-corrected, TZP-corrected and NZP-corrected RV time series. The weighted mean values of the non-corrected, TZP-corrected and NZP-corrected data are 3.25~m/s, 3.23~m/s and 3.09~m/s, respectively. This shows that the TZP correction indeed improves the instrument performance, although this improvement is small compared to the NZP correction, which provides better results. We note, however, that both correction methods work particularly well for those stars that have rather low RV standard deviations. This is to be expected as in these stars instrumental effects are not masked by larger intrinsic RV variations.

Altogether, the close relation between the theoretical TZP and the observed NZP strongly suggests that the latter are mainly driven by the changes in the environment. While the NZP correction yields better results, the TZP provides a description of the instrument, independently of the stars observed during the night. 
We emphasize that the NZP strategy is based in the observation of stellar targets with similar spectral content -- all stars used in the NZP calculation are M-stars. {Given the nature of the CARMENES survey, this strategy has shown to be succesful for reaching precision stellar RVs. However, our results show that the long-term variations observed in the NZPs could be mitigated by providing an additional thermal control loop in the optical bench. We acknowledge that such intervention is out of the scope of this research, but we highlight the capabilities of \texttt{moes} to be used a diagnose tool for complex astronomical instrumentation.}
This shows the strength of our physical modeling approach to improve the performance of precision RV instruments, just using the daily calibrations and the environmental data of the spectrograph.

Our TZP method could become a valuable tool for precision RV instruments. {Its capabilities go from optical design to building up wavelength solutions to measure precision stellar RVs. By adding the environmental dependancy of the optical path of light rays through the instrument optics, we can distinguish between mechanically-driven changes in the spectra with the ones induced by the smooth changes in the environment, providing a useful diagnose tool to identify optical misalignments in echelle spectrographs.}
Further improvements in our software consider the calibration of stellar spectra using the physical modeling of the WLS, including the change of the $\delta f_n$ values in the WLS. With this strategy one could characterize how much of the NZP observed in the CARMENES VIS data is influenced by the environment, or even more, calibrate the data in such a way that no NZP is required to optimize the instrument performance.

\section{Conclusion}\label{sec:Conclusion}

We have built a ray tracing model of the CARMENES VIS spectrograph to create a physical WLS. The inclusion of aberration terms enables us to reach the sub-pixel accuracy required to measure precision RVs. A combination of SA and NS algorithms allows us to calculate the instrumental parameters that best fit the calibration data and to study existing correlations between the instrumental parameters.

Using our ray tracing WLS, we have calculated the change in the relative positions of the science and calibration fibers as a function of temperature and pressure, with the goal of calculating RV systematics arising from changes in the spectrograph environment. These environmentally-driven RV zero points, which we label as TZP, follow the same trend as the observed NZP and the measured temperature gradient inside the spectrograph.
Our results and analysis support the hypothesis that the temperature gradient variations, coupled with pressure changes in the CARMENES VIS spectrograph, are the main cause for the observed NZPs in the stellar RV time series.

Further improvements consider the calculation of RVs using the {\tt moes} WLS, and the extension of {\tt moes} to other precision stellar RV facilities. The use of ray tracing models to predict RV systematics in precision RV instruments will become a valuable tool for the community, as it will allow to create new observing and calibration strategies, by providing a deeper understanding of the instrument systematics. 

\section*{Data Availability}

All the data and codes that support the findings of this study are openly available in the GitHub repository at \url{https://github.com/mtalapinto/moes/}. The WLS data for the CARMENES VIS spectrograph extracted from the \texttt{caracal} pipeline are stored at \url{https://github.com/mtalapinto/moes/tree/main/carmenes/data/vis_ws_timeseries} and the CARMENES VIS RVs used in this study can be found in \url{https://github.com/mtalapinto/moes/tree/main/carmenes/caracal_rvs}. 
The instrumental parameters of the \texttt{moes} WLS and the chromatic and aberration coefficients time series can be found in \url{https://github.com/mtalapinto/moes/tree/main/carmenes/data}, where we have also included a folder with the TZP values calculated for all the stars considered in our study.

\section*{Acknowledgements}
CARMENES is an instrument at the Centro Astronómico Hispano-Alemán at Calar Alto (CAHA, Almería, Spain). CARMENES is funded by the German Max-Planck-Gesellschaft (MPG), the Spanish Consejo Superior de Investigaciones Científicas (CSIC), 
the European Union through FEDER/ERF FICTS-2011-02 funds, and the members of the CARMENES Consortium (Max-Planck-Institut für Astronomie, Instituto de Astrofísica de Andalucía,
Landessternwarte Königstuhl, Institut de Ciències de l’Espai, Institut für Astrophysik Göttingen, Universidad Complutense de Madrid, Thüringer Landessternwarte Tautenburg, 
Instituto de Astrofísica de Canarias, Hamburger Sternwarte, Centro de Astrobiología and Centro Astronómico Hispano-Alemán), 
with additional contributions by the Spanish Ministry of Economy, the German Science Foundation through the Major Research Instrumentation Program and DFG Research Unit FOR2544 “Blue Planets around Red Stars”, the Klaus Tschira Stiftung, 
the states of Baden-Württemberg and Niedersachsen, and by the Junta de Andalucía. 

M.T.P.~acknowledges support by the International Max Planck Research School for Astronomy and Cosmic Physics at the University of Heidelberg, the Deutscher Akademischer Austauschdienst through Graduate Scholarship Program no. 91593263, and Agencia Nacional de Investigación y Desarrollo through the FONDECYT Postdoctoral Fellowship no. 3210253.

We thank Dr.~Florian Bauer, who played a crucial role in the understanding of the wavelength solution for the CARMENES RVs.


\bibliography{article}   
\bibliographystyle{spiejour}   


\appendix

\section{Thermal expansion modeling}\label{app:environment}

Materials tend to change their geometry when they are exposed to temperature changes. 
In \texttt{moes} we implement a linear expansion model of the form
\begin{equation}\label{eq:cte}
    \Delta L = \alpha_L L \,\Delta T,
\end{equation}
where $\alpha_L$ is the Coefficient of Thermal Expansion (CTE), $L$ is the length of the material and $\Delta T$ is the difference in temperature. This approximation is suitable only if $\alpha_L$ does not change much as a function of $\Delta T$ and the fractional length $\Delta L/L \ll 1$. In \texttt{moes} the parameters affected by thermal expansion are: the curvature radii and thicknesses of lenses, the curvature of mirrors, the orientation angles of optical elements, and the distances between the optical elements.


\vspace{2ex}\noindent{Marcelo Tala Pinto} is an Astronomer at Universidad Adolfo Ibañez. He received his BS and MS degrees in physics from the Pontificia Universidad Católica de Chile in 2013 and 2015, respectively, and his Dr. rer. nat. degree from the Karl-Ruprecht-Universität Heidelberg in 2019.  His current research interests include exoplanet detection and characterization, echelle spectroscopy, optical fibers, and optoelectronic systems.

\vspace{1ex}
\noindent Biographies and photographs of the other authors are not available.

\listoffigures
\listoftables

\end{spacing}
\end{document}